%% file: _index.tex
\documentclass[preprint]{vgtc}               

\ifpdf
  \pdfoutput=1\relax                   
  \usepackage{graphicx}                
  \DeclareGraphicsExtensions{.pdf,.png,.jpg,.jpeg} 
\else
  \usepackage{graphicx}                
  \DeclareGraphicsExtensions{.eps}     
\fi%

\graphicspath{{figures/}{pictures/}{images/}{./}} 

\usepackage{microtype}                 
\PassOptionsToPackage{warn}{textcomp}  
\usepackage{textcomp}                  
\usepackage{mathptmx}                  
\usepackage{times}                     
\usepackage{cite}                      
\usepackage{tabu}                      
\usepackage{booktabs}                  

\onlineid{0}

\vgtccategory{Research}

\vgtcinsertpkg

\title{Beyond English: Centering Multilingualism in Data Visualization}

\author{No\"{e}lle Rakotondravony\thanks{e-mail: ntrakotondravony@wpi.edu} \footnotemark[4]\\ %
        \scriptsize University of Konstanz, \scriptsize University of Antananarivo, \\ %
        \scriptsize Worcester Polytechnic Institute
\and Priya Dhawka \thanks{e-mail: dhawkapriya@gmail.com} \footnotemark[4]\\ %
     \scriptsize University of Calgary\\ %
     \scriptsize University of Washington %
\and Melanie Bancilhon\thanks{e-mail: mbancilhon@wustl.edu} \thanks{The authors contributed equally to this work.}\\ %
     {\scriptsize Washington University in St. Louis}}
\teaser{
}

\abstract{
Information visualization and natural language are intricately linked. However, the majority of research and relevant work in information and data visualization (and human-computer interaction) involve English-speaking populations as both researchers and participants, are published in English, and are presented predominantly at English-speaking venues. 
Although several solutions can be proposed such as translating English texts in visualization to other languages, there is little research that looks at the intersection of data visualization and different languages, and the implications that current visualization practices have on non-English speaking communities. 
In this position paper, we argue that linguistically diverse communities abound beyond the English-speaking world and offer a richness of experiences for the visualization research community to engage with. 
Through a case study of how two non-English languages interplay with data visualization reasoning in Madagascar, we describe how monolingualism in data visualization impacts the experiences of underrepresented populations and emphasize potential harm to these communities.  
Lastly, we raise several questions towards advocating for more inclusive visualization practices that center the diverse experiences of linguistically underrepresented populations.} 

\CCScatlist{
  \CCScatTwelve{Human-centered computing}{Visu\-al\-iza\-tion}{Visualization design and evaluation methods}{}
}

\usepackage{hyperref}
\usepackage{todonotes}

\usepackage{xspace}
\newcommand{\etal}{\emph{et al.}\@\xspace}

\usepackage{enumitem} 

\begin{document}
\maketitle

\input{_main}

\bibliographystyle{abbrv-doi}

\bibliography{_vis4good2023}
\end{document}

%% file: _main.tex
\section{Introduction}
Visualizing data is fundamentally about communicating information to viewers. 
To this end, designers and researchers often incorporate elements such as descriptive text, and symbols to more effectively communicate the information being represented. 
For instance, including human-readable text in natural language to explain a data visualization is a nearly universal practice in visualization research and design communities~\cite{hearstshow}. 
With the majority of human-computer interaction researchers, including those working in information visualization, being based in English-speaking WEIRD countries~\cite{linxen2021weird}, English is used extensively as the language of choice in data visualization research. 
However, researchers and designers outside of predominantly English-speaking countries have long been using data visualizations to communicate information across the globe, in a variety of non-English speaking cultures. 

As online and hybrid conferences have widened access to the visualization research community in recent years, it has become more apparent that existing practices within visualization research and design (such as monolingualism) may contribute towards excluding underrepresented and under-served groups from the visualization research community. 
Despite ongoing efforts to foreground inclusion and equity in our research and design practices as a community, we have often neglected language as a factor that may influence who gets to participate in and benefit from visualization research. 

Hence, in this paper, we ask: 

\begin{itemize}[noitemsep]
    \item What is the impact of monolingual, English-speaking visualization research practices on underrepresented communities, such as individuals from non-English speaking cultures?
    \item How do monolingual visualization practices further the exclusion of already underrepresented and under-served communities in visualization research?
\end{itemize}

We focus specifically on language used for descriptive text, human-readable captions, and communication with data visualizations. 
We illustrate our discussion with a case study on the use of two non-English languages and their interactions with data visualization in Madagascar where multilingualism is a result of colonization, imposing distinct languages for colloquial communications and for formal instruction. 
We argue that simple measures, such as translating English text in data visualizations to local languages, are accessible but fundamentally ignore the broader issue at hand --- that monolingual research practices hide how the lived experiences of viewers may influence the ways in which they interact with visualizations. 
We end with a call to action to the visualization research community to examine our current practices for exclusionary effects and provide suggestions for potential research directions to increase multilingualism in visualization research. 




\section{Related Work}
We describe previous research within information visualization and human-computer interaction (HCI) with a focus on critical data visualization and HCI.

\subsection{Critical Data Visualization and HCI}
Scholars working in critical data visualization and HCI advocate for equitable research practices when working with marginalized and underrepresented populations. 
Within the context of information visualization, D\"{o}rk \etal~\cite{doerk2013infovis} outline a number of ways in which researchers can question how their values and assumptions pervade their existing research practices. 
Specifically, they suggest research practices that prioritize disclosure, contingency, plurality and empowerment~\cite{doerk2013infovis}. 

On the HCI side, D'Ignazio and Klein propose a data feminism framework for researchers and practitioners to question and disrupt inequitable research practices when working with marginalized populations~\cite{datafem}. 
They urge researchers to challenge assumptions about gender binaries, power structures, and the objectivity of data while incorporating context and the diverse lived experiences of the people behind the data~\cite{datafem}. 
In their call to action to the HCI community, Ogbonnaya-Ogburu \etal propose ways to adapt critical race theory learnings to HCI research. 
They elucidate how racism is pervasive in existing research practices, and advocate for race-conscious research practices~\cite{Ogbonnaya2020}. 

Work within critical visualization has also started considering the ethics of unquestioned research practices on audiences. 
For instance, Correll highlights potential side-effects of standard visualizations such as viewers feeling alienated from the people whose data is being represented~\cite{correll2019ethical}.  
Meanwhile, in their study of attitudes towards data in rural Pennsylvania, Peck \etal~\cite{peck2019data} found that one-size-fits-all approaches in visualization research tend to neglect certain demographic populations and that individuals' personal lived experiences strongly influence how they relate to data visualizations.

\section{The Integration of Visualization \& Text Should Consider Language} \label{sec:vis-text-language}

When it comes to information communication, the language surrounding visualization has been shown to be equally important as the chart, if not more. 
Several studies have shown that integrating text and visualization can improve recall~\cite{borkin2015beyond}. People have also reported to prefer formats that contain text~\cite{stokes2022more,bancilhon2023communicating}. The integration of text and charts has been examined across a number of applications, including high-stake tasks such as Bayesian reasoning, which is notoriously difficult for most people. 

Bancilhon \etal found evidence that integrating text and visualization reduced cognitive load when solving a Bayesian task compared to when showing either representation alone~\cite{bancilhon2023combining}. 
However, Ottley \etal has shown that text, visualization and a combination of the two formats elicit the same speed and accuracy in Bayesian tasks~\cite{ottley2015improving}. 
When examining eye tracking patterns during a study task, they found that participants likely identified critical information more effectively using visualization, but extracted information more effectively from text ~\cite{ottley2019curious}. 
These mixed findings show that the integration of text and chart is not straightforward. While several other applications have shown that integration techniques can improve recall and accuracy~\cite{zhi2019linking}, techniques such as linking and hovering have shown no measurable improvement in the context of Bayesian reasoning~\cite{mosca2021does}. 
More research needs to be conducted to examine the best techniques to integrate the two such that their respective benefits are optimized. 

While format and layout can impact reasoning when integrating text and charts, semantics can also play a role. Stokes \etal found that texts that describe statistical or relational components of a chart lead to more takeaways than texts that describe elemental or other components~\cite{stokes2022striking}. 
Other studies have shown that the title of a visualization can influence its interpretation more than the visualization itself~\cite{kong2018frames}. 

A majority of the studies examining the effects of integrating chart, text, and language have been conducted in English. 
A study by Rakotondravony \etal showed an example of how the verbalization of quantitative  probability through visualizations can vary across French, Arabic, English, German, and Mandarin~\cite{rakotondravony2022probablement}. 
However, further studies are still needed to build evidence on how text interacts with visualization in under-represented languages. 
Moreover, when evaluating reasoning using charts, there needs to be important considerations for individual differences in visual literacy, which can be influenced by language and culture. There lacks research on developing valid and inclusive visualization literacy assessments. Pandey \etal, who developed a shortened version of the visualization literacy test, posit that more research needs to be done to adopt the MINI-VLAT in different languages~\cite{pandey2023mini}.


In emerging fields such as AI and NLP, there is an increasing amount of research on text and visualization integration. 
Several researchers have investigated ways to assist the visually impaired or people with low visualization literacy via techniques such as chart summarization~\cite{obeid2020chart}. 
Other researchers like Shen \etal  have examined how to automate chart generation using natural language and conducted a survey of natural language interfaces for data visualization~\cite{shen2022towards}. 
The lack of consideration for language persists across these fields, where most of the development and evaluation is done in English.


\section{Madagascar: A case study of language, data visualization, in non-English }\label{sec:case-madagascar}

In this section, we illustrate the interplay between language and data visualization in a non-English speaking context.
We describe how the affordance of language can impact the expressiveness of data visualizations, and challenge the different sub-areas of visualization research. 
The case of Madagascar that we highlight is likely common to countries and communities speaking more than one official language, and for which the spoken colloquial languages are different from that of instruction.
In Madagascar, Malagasy is the country's national language. 
It is spoken and understood overall in its different dialectical variations~\cite{bouwer2005towards}.
In the post colonial era, French was instituted as language of instruction, and was reaffirmed as a \textit{de jure} official language by the country's constitution in 2007. 
In public schools through grade five, Malagasy is the language of instruction for all subjects, whereas in high school it is for the subjects of history and Malagasy language only. 
Instruction for science and the other subjects is delivered in French. 

While the majority of people have been in contact with the French language in their primary years, most rarely, if not never, practice it after school.
As of 2022, it is estimated that $26.5\%$ of Malagasy population older than 10 years old speak French at least at a colloquial fluency~\cite{marcoux2022estimation}.
Beyond being the predominant language of education in primary and higher education, French is also used and referenced for most technology-related topics on the island\footnote{\href{https://observatoire.francophonie.org/qui-apprend-le-francais-dans-le-monde/le-francais-langue-denseignement/}{Le Français langue d'enseignement}, accessed: 2023-07-10}. 

\subsection{Communicating With Data Visualization}
\begin{figure}[ht]
    \centering
    \includegraphics[width= \linewidth]{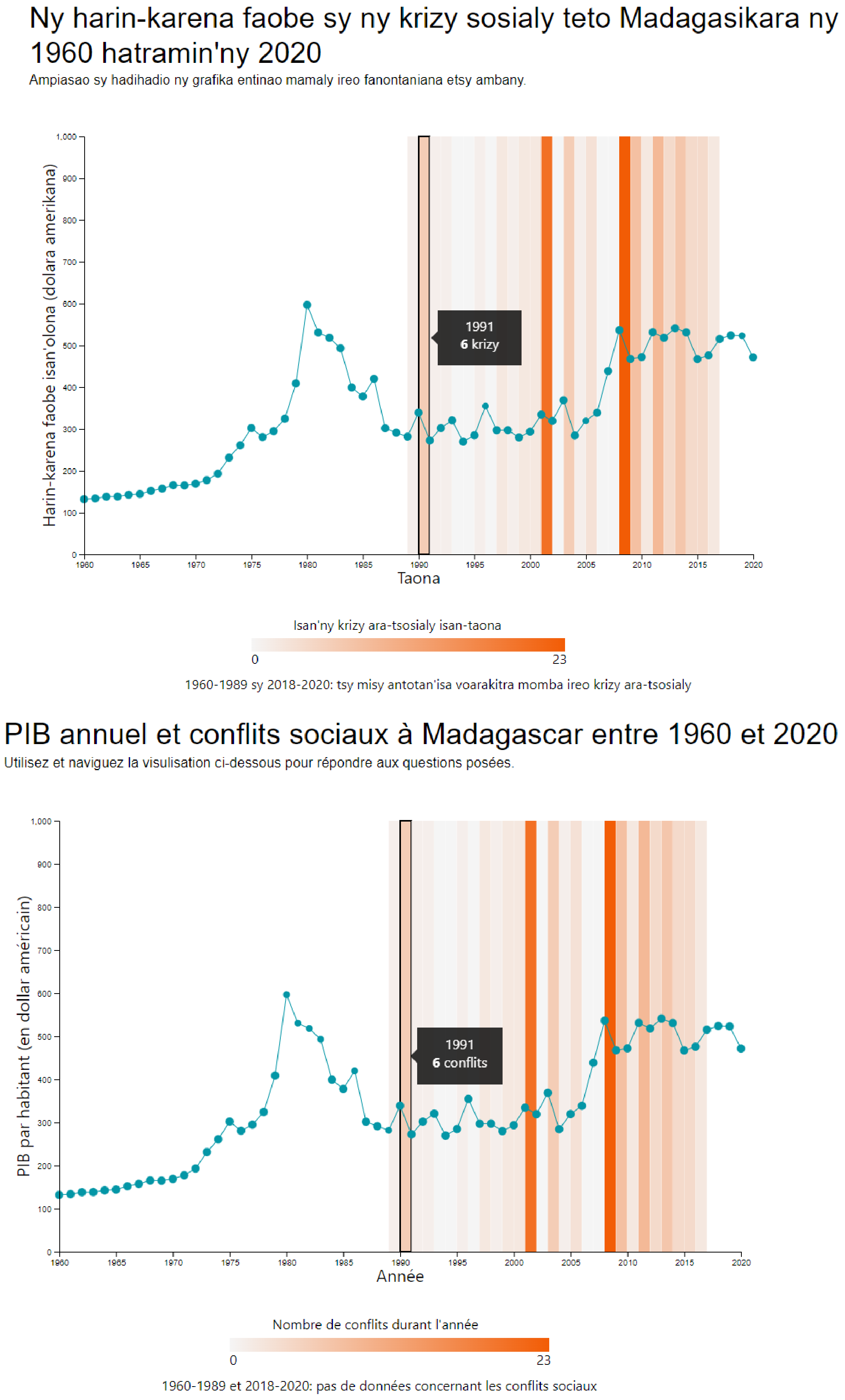}
    \caption{Stimulus used in Andrianarivony \etal's exploratory study investigating how bilingual audiences use their native and secondary languages with visualizations in Madagascar ~\cite{andrianarivony2022investigating}. Image courtesy of the authors.}
    \label{fig:poster-stimulus}
\end{figure}
Data visualization today is about communicating (quantitative) information through static or interactive visuals and graphs. 
In line with existing observations of the performance of audiences from bilingual settings in mathematical reasoning~\cite{baker2011foundations, unesco2016understand} and in HCI ~\cite{evers1998cross, onibere2001human}, the prevalence of French in Madagascar, especially in science education and technology, can influence the statistical or quantitative reasoning with technological instruments, such as interactive interfaces, in Madagascar even among an highly educated audience.

In an exploratory study of how French and Malagasy are used to discuss data and data visualization in Madagascar, Andrianarivony \etal ~\cite{andrianarivony2022investigating} asked native speakers to describe their approach for completing basic visualization reading tasks such as finding a specific value, and to verbalize their takeaways from a data visualization in either Malagasy or French only.
The stimulus combines a line chart and a bar chart depicting the evolution of Madagascar's GDP and the recorded number of social conflicts in the country. 
Participants in the study hold at least a higher education degree and were fluent in both languages. 

Results highlighted the easiness of using French for participants to talk about how they interact with the charts, explicitly naming the different elements of the visuals (such as \textit{barchart, linechart}, ...), the actions that they took to explore it (such as \textit{clicking}, \textit{zooming}, ...), and their note on the data (such as \textit{increase}, \textit{decrease}, ...). 
On the other hand, participants in Malagasy showed difficulty to verbalize their analysis of both the chart and its underlying data, essentially in finding the necessary terminologies which they often already know in French. 
While the knowledge of French appears to be helpful, the lack of linguistic tools for verbalizing data and visualizations in their native language excludes approximately $74\%$ of the Malagasy population who do not speak French in Madagascar from benefiting from visual and graph-based communications. 
It also reflects the understudy of non-English languages in visualization and interaction studies.


\subsection{Data Visualization Literacy}
Beyond challenging the use of data visualization for communication, a lack of multilingualism in research also directly impacts visualization literacy. 
As visualization literacy gains interest among researchers, visualization literacy assessment tests constitute useful instruments to help develop future design and methodologies that help audiences better read and make sense of data visualizations.
However, most approaches in the research literature are in English, and assume underlying characteristics of the test takers that are not inclusive, especially for those from non-WEIRD contexts. 

For the majority of potential visualization users in Madagascar, knowledge of English constitutes a limitation to access the off-the-shelf assessment tests. 
Test items extensively refer to data, charts elements, and the use of statistical analyses.
They require fluency in charts naming conventions, mathematics, and data reasoning, and their contents are sourced from materials that do not necessarily align with other educational systems and curricula outside that of the country in which they were developed.
For example, for VLAT~\cite{lee2016vlat}, test-takers are expected to possess graph reading and interpretation skills comparable to the K-12 curriculum. 
In Madagascar, students are extensively exposed to graph reading at a much later stage in their course of studies.
This difference in curriculum required for data visualization literacy and tests might therefore impact participants' performance, potentially leading to incorrect conclusions.

While translation from English constitutes a good alternative to the lack of multilingualism of data visualization, more aspects of data visualizations beyond text translation still need to be considered and culturally adapted to ensure adequate evaluations.
As Peck \etal ~\cite{peck2019data}'s study showed, the way individuals relate to data and visualizations influence their engagement with a visual representation. 
While we hypothesize that adapting the western-centered visualization literacy test items to local contexts can increase test takers' interest in the data, more evidence and studies are still needed to understand how such engagement may impact their visualization literacy scores.



In this illustrative case of Madagascar, we argued how a lack of multilingualism in data and information visualization research can exclude underrepresented communities from participating in research studies. 
Similar to Madagascar, the described duality in language is shared among many other countries and post colonial countries, where the inherited or imposed language of instruction does not align with the colloquial language, creating among other challenges, the unspoken divide between who can benefit from data and information visualization and who cannot.
While we acknowledge that this issue is at core systemic, we argue that the information visualization research community can actively contribute to closing the gap.

\section{What is the impact of monolingualism and what can we do about it? }

Data visualization and text, hence language, are intrinsically linked~\cite{hearstshow}. 
The interplay of visualizations along with integrated text are many times used to communicate critical information that can impact decision-making across different levels, and to promote social good.
Reflecting on the case in \autoref{sec:case-madagascar}, a direct harm that the lack of multilingualism in our research practices can pose is the exclusion of marginalized populations.

\subsection{The Exclusion of Marginalized Populations}

Data about marginalized and underrepresented populations is often incomplete, does not exist or requires additional care for the privacy and anonymity of the people in the dataset. 
Available data often focus on their experiences of exclusion and suffering. 
For instance, anthropographics (human-shaped visualizations) of marginalized and underrepresented populations are widely used within English-based information visualization research in charitable giving settings or in ``humanitarian" visualizations~\cite{boy2017showing, Morais2021}. 
However, marginalized and underrepresented populations are \textbf{rarely} involved in visualization research projects, even the ones directly impacting them~\cite{dhawka2022representing}. 
In the case of English-based anthropographics, individuals belonging to the populations being represented (refugees, victims of violence) may not even be able to meaningfully interact with these visualizations due to language barriers. 
Hence, it is essential that visualization research projects about marginalized and underrepresented populations find creative ways to involve said groups of people in the research process~\cite{dhawka2023}.

Moreover, the groups that are historically underrepresented from research are mostly from non-WEIRD countries or countries with under-developed resources for research. 
When their particular needs are not reflected in research carried out in WEIRD countries for use globally, any claim for \textit{generalizability} of results perpetuates the imposition of culturally non-adapted approaches, misleadingly labelled as development efforts.
For instance, as data visualization literacy is a fundamental skill and the basis of the development of methods for teaching data visualization, the exclusion of marginalized groups' lived experiences can result in materials that are inadequate for local education systems beyond WEIRD countries. 
This inadequacy is often more prevalent in countries where the languages of instructions are not the native and spoken languages for communication~\cite{unesco2016understand}.







\subsection{Call to Action}
We propose actions for the information visualization community towards increasing the involvement of linguistically underrepresented populations in research. 
We emphasize that our propositions are preliminary ideas intended to start a community conversation around our existing research practices, and that several of these ideas will require additional long-term collective thinking and active support before implementation.

\paragraph{\textbf{Participatory Design and Visualization}}
Although information visualization research projects lean heavy on the interaction side, as a community, we often engage with research participants through short studies and in laboratory settings. 
This is in part due to the nature of our field of research being largely quantitative. 
However, as rich, qualitative studies become more common in visualization research, we can consider investigating how the lived experiences of viewers influence how they interact with data visualizations. 
For instance, future work can explore the (side) effects of using English in visualizations for underrepresented communities by directly involving said communities in visualization design workshops and investigating their interactions with multilingual visualizations. 
Additional research opportunities could focus on data visualization literacy in multiple languages as well as in countries with specific linguistic cultures.
 
\paragraph{\textbf{Leveraging Emerging ML Technologies}}
With the rapid development of ML technologies, the information visualization has already started speculating about the potential of incorporating these technologies within existing visualization workflows. 
In terms of natural language, access to LLMs in a variety of languages, faster and on-demand language translation services can support linguistically underrepresented populations who may not have access to human translators or where there are few alternatives to English-based visualizations. 
We imagine research opportunities around automated visualization tools with LLM-powered translation features. 
Similarly, we imagine visualization design and literacy tools that provide human-readable descriptions and captions in languages other than English, that combine multiple languages in a single interface as well as ones that include expertise from experts outside of data visualization (such as linguists) to address the challenges of learning additional languages. 

\paragraph{\textbf{Expanding Access to the Visualization Community}}
The COVID-19 pandemic saw the visualization community adapt to remote and hybrid conferences, which opened up access to communities who previously could not attend these conferences due to a number of financial, geographical and border-crossing limitations. 
Despite the many lessons learnt running successful remote and hybrid conferences during this time, a return to \textit{``normal''}, in-person only conferences can erase much of the progress around conference accessibility made during the pandemic. 

We envision smaller satellite conferences and remote workshops, in local languages, in under-served countries to increase researcher diversity within information visualization research and to support languistic diversity. 
For instance, within the HCI community, several geographically-bound conferences such as NordicCHI and AfriCHI already take place throughout the year. 
Although the information visualization community is smaller, we imagine a future with more remote opportunities to be in community with visualization researchers from underrepresented countries through workshop or paper tracks for specific geographic regions and dedicated mentorship networks. 

Additionally, there has been recent interest within the information visualization community to examine our publication practices. For instance, Hao and colleagues~\cite{hao2022thirty} examined 32 years of publications and citations at IEEE VIS venues and found that the majority of cited and referenced papers were limited to the same subfields of Computer Science. Building on this work, as a community, we may reflect on our citational practices, as Kumar and Karusala~\cite{kumar2021} urge HCI researchers: Why do we cite the way we do in information visualization research? Who do we cite (and not cite) and why? Where does the majority of our knowledge in information visualization research come from and why? 





\paragraph{}
Lastly, several challenges may arise when implementing actions towards inclusive visualization research that go beyond simple translations.
For instance, collaborations can be costly in both material and intellectual resources. Conducting studies online requires participants to have access to the necessary technologies. Some training and study configurations require in-person collaboration for which long distance, international travels are necessary. 
International mobility, beyond issues related to its environmental sustainability, comes with many barriers to the collaborations between researchers across countries but may particularly impact those from the Global South with limited border-crossing privileges.

Moreover, conducting data visualization research in languages other than English requires peer-reviewers who understand the language. 
While there are certainly experienced reviewers fluent in non-English languages, the community still needs to reflect on and discuss the best practices for including under-represented languages in research while being mindful of the potential of creating isolated findings that cannot be communicated widely. 

Despite these challenges, we strongly advocate for the visualization community to start this important conversation on increasing linguistic diversity in our research practices.

\section{Conclusion}
We discussed how monolingual practices in the data visualization community creates a divide between who gets included in or benefits from visualization research. 
We emphasize the importance of text and language in data visualization by highlighting how their interplay helps in promoting the access to existing data visualizations, and the inclusion of traditionally underrepresented populations in data visualization research. 
Reflecting from an exploratory study in Madagascar, where the lack of linguistic tools for data and visualizations impacts how people interact with charts and verbalize their takeaways, we highlight the potential harms that the exclusion of under-studied languages can cause --- especially when communicating data about and to marginalized and underrepresented populations.
We conclude this position paper by calling the community's attention to several discussion avenues towards increasing the involvement of underrepresented populations in visualization research through efforts such as participatory design, leveraging emerging ML technologies and expanding global access to the visualization research community. 
Our hope with this work is that data visualizations can be created, designed, used and accessed by all, especially those who make up the majority of the global population.